\documentclass[12pt]{article} 

\usepackage{amssymb} 
\usepackage{amsmath}
\usepackage{amsfonts}

\newcommand{\R}{\mathbb{R}}
\newcommand{\C}{\mathbb{C}}

\newcommand{\be}{\begin{equation}}
\newcommand{\bea}{\begin{eqnarray}}
\newcommand{\eea}{\end{eqnarray}}
\newcommand{\nn}{\nonumber}
\newcommand{\kt}{\rangle}
\newcommand{\br}{\langle}

\newcommand{\ed}{\end{document}}
\newcommand{\bbr}{\br\!\br}
\newcommand{\kkt}{\kt\!\kt}
\newcommand{\cbr}{(\!(}
\newcommand{\ckt}{)\!)}

\begin{document}

\title{Hilbert Space Structures on the Solution Space of Klein-Gordon Type Evolution Equations}
\author{Ali Mostafazadeh\thanks{E-mail address: amostafazadeh@ku.edu.tr}\\ \\
Department of Mathematics, Ko\c{c} University,\\
Rumelifeneri Yolu, 34450 Sariyer, Istanbul, Turkey}
\date{ }
\maketitle

\begin{abstract}
We use the theory of pseudo-Hermitian operators to address the problem of the construction and classification of positive-definite invariant inner-products on the space of solutions of a Klein-Gordon type evolution equation. This involves dealing with the peculiarities of formulating a unitary quantum dynamics in a Hilbert space with a time-dependent inner product. We apply our general results to obtain possible Hilbert space structures on the solution space of the equation of motion for
a classical simple harmonic oscillator, a free Klein-Gordon equation, and the Wheeler-DeWitt equation for the FRW-massive-real-scalar-field models.
\end{abstract}

\baselineskip=24pt

\section{Introduction}

The problem of finding a consistent probability interpretation for Klein-Gordon fields is almost as old as quantum mechanics. It was this problem that led Dirac to consider his celebrated field equation for the spin half particles. Although the discovery of the Dirac equation had an overwhelming impact on the development of modern physics and mathematics, it did not solve the problem of the probability interpretation for Klein-Gordon fields. This problem was actually put aside after Dirac's discovery of the method of second quantization. The first quantized scalar fields have limited applications in relativistic quantum mechanics. They are also plagued by the so-called Klein paradox whose resolution calls for a second quantized field theoretic treatment. These are the main reasons why the issue of the probability interpretation for Klein-Gordon fields did not attract much attention in the 1930s-1950s. 

The situation changed drastically in the 1960's by the advent of quantum cosmology and the Wheeler-DeWitt equation. This is a Klein-Gordon type field equation whose solution is identified with the wave function of the universe. The interest in finding a consistent probability interpretation for Klein-Gordon fields was therefore revived as it constituted a fundamental obstacle in trying to make sense of quantum cosmology. Unlike the case of the Klein-Gordon fields, the method of second quantization and Dirac's trick of considering an associated first order field equation cannot be used to escape dealing with this problem in quantum cosmology \cite{kuchar,isham,wiltshire,dereli,carlip}.  As a complete solution was out of reach, various partial solutions or rather attempts to avoid this problem were considered in the literature. These are either based on the invariant but indefinite Klein-Gordon inner-product which was originally proposed in DeWitt's pioneering article \cite{bryce-1} and further developed by others \cite{vilenkin,wald}, or use the Wheeler-DeWitt field to define a conditional probability 
\cite{page-w,page}. These attempts have however been subject to controversy and criticism; see for example
\cite{kuchar,isham}. 

In mathematical terms, devising a consistent probability interpretation for the solutions of a field equation is equivalent to finding an invariant positive-definite inner product on the space of solutions of this equation. Here the invariance of the inner product means that the inner product of two solutions is independent of time. This in turn brings up the notorious `problem of time' \cite{kuchar,isham,page,wiltshire}. Disregarding this problem, i.e., assuming that a time-variable is selected, reduces the goal of finding a probability interpretation for the Wheeler-DeWitt field to the construction of an invariant positive-definite inner product on the solution space of the Wheeler-DeWitt equation, i.e., to address the so-called Hilbert space problem \cite{kuchar}. The first successful attempt in this direction is due to Woodard \cite{woodard} who made use of the idea of gauge-fixing the Wheeler-DeWitt symmetry. More recently, a closely related approach, namely the method of refined algebraic quantization and group averaging, has been developed \cite{marolf}. This
approach is technically involved and so far could only be employed in the study of some special models. For a related development, see \cite{gp}.

The purpose of this article is to use the recent results on pseudo-Hermitian Hamiltonians 
\cite{p1,p2,p3,p4,p8} to address the problem of the existence/construction and uniqueness/classification of the invariant positive-definite inner products on the space of solutions of a Klein-Gordon type field equation. 

The organization of the article is as follows. In Section~2, we present a brief review of the basic properties of pseudo-Hermitian Hamiltonians. In Section~3, we discuss a formulation of quantum mechanics for a system with a time-dependent Hilbert space. In Section~4, we consider the application of the theory of pseudo-Hermitian operators to a class of generic Klein-Gordon type field equations. In Sections 5-8, we apply our general results to a classical harmonic oscillator, the free Klein-Gordon equation, and the Wheeler-DeWitt equation for the FRW-massive-real-scalar-field models, respectively. Finally in Section~9, we present a summary of our findings and discuss their implications.

\section{Pseudo-Hermitian Hamiltonians}

We begin our survey of the properties of pseudo-Hermitian operators by recalling that given a linear, Hermitian, invertible operator $\eta:{\cal H}\to{\cal H}$ acting in a
Hilbert space ${\cal H}$ with inner product $\br~|~\kt$, the function 
$\bbr~|~\kkt_\eta:{\cal H}^2\to\C$ defined by
	\be
	\bbr\psi_1|\psi_2\kkt_\eta:=\br\psi_1|\eta\psi_2\kt
	\label{ps-inner}
	\end{equation}
is a possibly indefinite inner product on ${\cal H}$. For brevity we shall call the latter a {\em pseudo-inner product}. In precise terms, a pseudo-inner product $\bbr~|~\kkt$ on a vector space ${\cal H}$ is 
a quadratic form (i.e., it maps any two elements $\psi_1$ and $\psi_2$ of ${\cal H}$ to a complex number) with the following properties \cite{kato}.
	\begin{enumerate}
	\item It is nondegenerate, i.e., if for all $\psi_1\in{\cal H}$, $\bbr\psi_1|\psi_2\kkt=0$, then 	$\psi_2=0$;
	\item It is Hermitian, i.e., for all $\psi_1,\psi_2\in{\cal H}$, $\bbr\psi_1|\psi_2\kkt^*=
	\bbr\psi_2|\psi_1\kkt$, where $*$ denotes complex-conjugation;
	\item It is sesquilinear, i.e., for all $\psi_1,\psi_2,\psi_3\in{\cal H}$ and for all 
	$\alpha,\beta\in\C$,
	\[\bbr\psi_1|\alpha\psi_2+\beta\psi_3\kkt=\alpha\bbr\psi_1|\psi_2\kkt+\beta
	\bbr\psi_1|\psi_3\kkt.\]
	\end{enumerate}
It is not difficult to observe that any pseudo-inner product on a Hilbert space ${\cal H}$ is of the form $\bbr~|~\kkt_\eta$ for some linear, Hermitian, invertible operator $\eta:{\cal H}\to{\cal H}$.

A linear operator $H:{\cal H}\to{\cal H}$ is said to be {\em pseudo-Hermitian} \cite{p1} if there is a linear, Hermitian, invertible operator $\eta:{\cal H}\to{\cal H}$ such that
	\be
	H^\dagger=\eta H\eta^{-1}.
	\label{pseudo}
	\end{equation}
A pseudo-Hermitian operator together with a given operator $\eta$ satisfying~(\ref{pseudo}) is said to be {\em $\eta$-pseudo-Hermitian}.

The basic properties of pseudo-Hermitian operators are the following \cite{p1,p2,p3,p8}.
	\begin{itemize}
	\item[] {\bf Theorem~I:} $H$ is $\eta$-pseudo-Hermitian if and only if it is Hermitian with respect to the pseudo-inner product $\bbr~|~\kkt_\eta$, i.e., for all $\psi_1,\psi_2\in{\cal H}$, $\bbr\psi_1|H\psi_2\kkt_\eta=\bbr H\psi_1|\psi_2\kkt_\eta.$
	\item[] {\bf Theorem~I$\!$I:} Let $H$ be the Hamiltonian of a quantum system and $\eta$ be a linear, Hermitian, invertible operator. Suppose that $\eta$ is time-independent, then $H$ is $\eta$-pseudo-Hermitian if and only if the pseudo-inner product $\bbr~|~\kkt_\eta$ is a dynamical invariant. That is given any two solutions $\psi_1(t)$ and 
$\psi_2(t)$ of the Schr\"odinger equation, 
	\be
	i\hbar\frac{d}{dt} \psi= H\psi,
	\label{sch-eq}
	\end{equation}
$\bbr\psi_1(t)|\psi_2(t)\kkt_\eta$ does not depend on time. If $\eta$ depends on time, the pseudo-Hermiticity of $H$ implies
	\be
	\frac{d}{dt}\,\bbr\psi_1(t)|\psi_2(t)\kkt_{\eta(t)}=
	\br\psi_1(t)|\frac{d\eta(t)}{dt}\:\psi_2(t)\kt=
	\bbr\psi_1(t)|\eta^{-1}(t)\:\frac{d\eta(t)}{dt}\:\psi_2(t)\kkt_{\eta(t)}.
	\label{t-dep}
	\end{equation}
\item[] {\bf Theorem~I$\!$I$\!$I:} Let $H$ be a diagonalizable Hamiltonian with a discrete spectrum. Then the following are equivalent. 
	\begin{itemize}
	\item[] 1.~The eigenvalues of $H$ are either real or come in complex-conjugate pairs. In this case 	we say that $H$ has a pseudo-real spectrum; 
	\item[] 2.~$H$ is pseudo-Hermitian; 
	\item[] 3.~$H$ admits an antilinear symmetry generated by an invertible antilinear operator ${\cal 	X}$, i.e., $[H,{\cal X}]=0$.
	\end{itemize}	
\item[] {\bf Theorem~I$\!$V:} Let $H$ be a diagonalizable Hamiltonian with a discrete spectrum.
Then the following are equivalent.
	\begin{itemize}
	\item[] 1.~$H$ has a real spectrum; 
	\item[] 2.~$H$ is $O^\dagger O$-pseudo-Hermitian for an invertible operator $O$. Equivalently,
	$H$ is $\eta$-pseudo-Hermitian for a positive operator $\eta$; 
	\item[] 3.~$H$ is related to a Hermitian operator by a similarity transformation. Following 	\cite{quasi}, we then call $H$ quasi-Hermitian; 
	\item[] 4.~$H$ is Hermitian with respect to a positive-definite inner product.
	\end{itemize}
\end{itemize}

As pointed out in \cite{p4,p8}, for a given pseudo-Hermitian diagonalizable Hamiltonian $H$ the linear, Hermitian, invertible operators $\eta$ that make $H$ $\eta$-pseudo-Hermitian are, up to the choice of the eigenbasis of $H$, classified by a set of signs $\sigma_{n_0} $; $\eta$  has the general form
	\be
	\eta=\sum_{n_0} \sigma_{n_0} |\phi_{n_0}\kt\br\phi_{n_0}|+\sum_{n+} 	(|\phi_{n+}+\kt\br\phi_{n-}|
	+|\phi_{n-}\kt\br\phi_{n+}|),
	\label{eta}
	\end{equation}
where $n_0$, $n+$ and $n-$ are respectively the spectral labels associated with the eigenvalues with zero, positive, and negative imaginary parts, and $|\phi_n\kt$, with $n=n_0, n+,n-$, are the eigenvectors of $H^\dagger$ that together with the eigenvectors $|\psi_n\kt$ of $H$ form a complete biorthonormal system, i.e., they satisfy
	\be
	\br\phi_m|\psi_n\kt=\delta_{mn},~~~~~~~~~~~~ \sum_n |\psi_n\kt\br\phi_n|=1.
	\label{bi}
	\end{equation}
In view of Eqs.~(\ref{eta}) and (\ref{bi}), $\bbr\psi_{n_0}|\psi_{n_0}\kkt_{\eta}=\sigma_{n_0}$, and $\bbr\psi_{n\pm}|\psi_{n\pm}\kkt_{\eta}=0$. Therefore, the eigenvectors with complex eigenvalues have zero pseudo-norm; they are null (or ghost) vectors. Furthermore, suppose that we adopt the inner product corresponding to the operator:
	\be
	\eta_+:=\sum_{n_0}|\phi_{n_0}\kt\br\phi_{n_0}|+\sum_{n+=n-}
	(|\phi_{n+}+\kt\br\phi_{n-}|+|\phi_{n-}\kt\br\phi_{n+}|),
	\label{eta+}
	\end{equation}
which is obtained by setting $\sigma_{n_0}=+$ for all $n_0$. Then the eigenvectors with real eigenvalue have positive norm (squared). In particular, if the spectrum is real, the inner product $\bbr~|~\kkt_{\eta_+}$ is positive-definite, \cite{ba}. This is precisely the inner product whose existence is ensured by Theorem I$\!$V. 

It is important to note that the operator $\eta_+$ and the corresponding invariant inner product, which makes the norm (squared) of the eigenvectors with a real eigenvalue positive, are only unique up to the choice of the biorthonormal basis system $\{|\psi_n\kt,|\phi_n\kt\}$. This means that choosing different biorthonormal systems, one obtains different $\eta_+$ and $\bbr~|~\kkt_{\eta_+}$. Under a change of the eigenbasis vectors $|\psi_n\kt\to|\tilde\psi_n\kt$, there is an invertible linear operator $A:{\cal H}\to{\cal H}$ such that $|\tilde\psi_n\kt=A^{-1}|\psi_n\kt$, $|\phi_n\kt\to |\tilde\phi_n\kt=A^{\dagger}|\phi_n\kt$, and consequently
	\be
	\eta_+\to\tilde\eta_+:=A^{\dagger}\eta_+ A. 
	\label{eta-trans}
	\end{equation}
Note that the operator $A$ maps an eigenvector to another eigenvector with the same eigenvalue. This suggests that $A$ generates a symmetry of the Hamiltonian, i.e.,
	\be
	[H,A]=0.
	\label{sym}
	\end{equation}
This can indeed be directly checked using the spectral resolution of the Hamiltonian:
	\be
	H=\sum_n E_n |\tilde\psi_n\kt\br\tilde\phi_n|=A^{-1}\sum_n E_n |\psi_n\kt\br\phi_n|A=
	A^{-1}HA.
	\label{spectral}
	\end{equation}
Therefore, the operator $\eta_+$ of Eq.~(\ref{eta+}) and the corresponding inner product $\bbr~|~\kkt_{\eta_+}$ are unique up to the symmetries of the Hamiltonian. Note that although we have shown the validity of the preceding statement in the absence of degenerate eigenvalues of $H$, we can repeat the same analysis and verify Eqs.~(\ref{eta-trans}) -- (\ref{spectral}) in the general case where some or all of the eigenvalues are degenerate.
	
We wish to conclude this section with the following remarks.
	\begin{enumerate}
	\item
The term `pseudo-Hermitian' was introduced in \cite{p1}. But it turns out that mathematicians \cite{ba} had developed similar concepts in the study of vector spaces with an indefinite metric, and Pauli \cite{pauli} had made use of these concepts in his study of a formulation of the quantum electrodynamics due to Dirac \cite{dirac}. Note however that there is an important difference between the approach pursued in the context of spaces with an indefinite metric (including Pauli's contribution) and the point of view adopted  in \cite{p1}. While in the former one considers a space with a given $\eta$, in the latter one formulates the concept of pseudo-Hermiticity without having to fix a specific $\eta$. In fact, as emphasized in \cite{p4} and discussed above, $\eta$ is not unique. In particular for a given $\eta$-pseudo-Hermitian Hamiltonian with a nonpositive $\eta$, if the spectrum of $H$ is real, one can construct another $\eta$ which is positive and use it to obtain a positive-definite inner product with respect to which $H$ is Hermitian. In fact, as discussed above and shown in \cite{p8} the most general positive-definite inner product with this property has the form $\bbr~~|~~\kkt_{\tilde\eta_+}$ where $\tilde\eta_+$ is given by (\ref{eta-trans}). This observation is the basic idea of the present article. 
	\item
Here and also in Refs.~\cite{p1,p2,p3,p4} we have given the relevant formulas for diagonalizable Hamiltonians with a discrete spectrum. In Ref.~\cite{p7}, we show how one can relax the diagonalizability condition. Also as noted in \cite{p3}, the presence of a continuous part of the spectrum does not cause any serious problems. If the spectrum is continuous, we treat the spectral label $n$ as a continuous variable, replace the summations with integrations, and change the Kronecker deltas to Dirac deltas.
	\end{enumerate}

\section{Unitary Evolutions in a Time-Dependent Hilbert Space}

Let $H(t)$ be a time-dependent pseudo-Hermitian Hamiltonian. Then in general the vectors $|\phi_n\kt$ and consequently $\eta_+$ and $\bbr~|~\kkt_{\eta_+}$ are also time-dependent. Therefore, $H(t)$ is Hermitian with respect to the inner product $\bbr~|~\kkt_{\eta_+}$, but the evolution is not unitary. In fact, the Hermiticity of $H(t)$ is neither necessary nor sufficient for the unitarity of the evolution.\footnote{The only exception to this statement is when at least one of the eigenvalues of the Hamiltonian is time-dependent, but all its eigenvectors are constant. The latter is precisely the condition of the exactness of the adiabatic approximation \cite{pra-97a,nova}.} This observation shows how having a time-dependent Hilbert space contradicts some of the established facts about ordinary quantum mechanics where the Hilbert space is stationary. It also raises the issue of the existence and uniqueness (or classification) of the inner products that make the evolution unitary. 

The following theorem provides a characterization of the inner products supporting a unitary evolution.
\begin{itemize}
\item[] {\bf Theorem~V:} Let $H:{\cal H}\to{\cal H}$ be a possibly time-dependent and non-Hermitian Hamiltonian acting in a Hilbert space ${\cal H}$. Then the evolution described by the Schr\"odinger equation (\ref{sch-eq}) is unitary if and only if ${\cal H}$ is endowed with the pseudo-inner product
	\be
	\bbr~|~\kkt_{\rm inv}=\br~|\eta_{\rm inv}|~\kt,
	\label{e=e}
	\end{equation}
where
	\be
	\eta_{\rm inv}:=U(t,t_0)^{-1\dagger}\eta_0 U(t,t_0)^{-1},
	\label{eta-inv}
	\end{equation}
$t_0$ is the initial time, $\eta_0:{\cal H}\to{\cal H}$ is a Hermitian, invertible, linear operator, 
	\be
	U(t,t_0):={\cal T}\, e^{-\frac{i}{\hbar}\int_{t_0}^t H(t')dt'}
	\label{evo}
	\end{equation}
is the evolution operator, and ${\cal T}$ is the time-ordering operator.
\item[] {\bf Proof:} First note that every pseudo-inner product can be written in the form~(\ref{e=e})
for some Hermitian, invertible, linear operator $\eta_{\rm inv}:{\cal H}\to{\cal H}$. Now, take two arbitrary solutions $\psi_1$ and $\psi_2$ of the Schr\"odinger equation (\ref{sch-eq}) and demand that 
$\bbr\psi_1|\psi_2\kkt_{\rm inv}$ is constant, i.e., for all $t\geq t_0$, $\br\psi_1(t)|\eta_{\rm inv}(t)|\psi_2(t)\kt=\br\psi_1(t_0)|\eta_{\rm inv}(t_0)|\psi_2(t)\kt$, alternatively $U(t,t_0)^\dagger\eta_{\rm inv}(t)U(t,t_0)=\eta_{\rm inv}(t_0)$. The latter equation is equivalent to
(\ref{eta-inv}) with $\eta_0=\eta_{\rm inv}(t_0)$.~~~$\square$
	\end{itemize}
In case of a time-independent Hamiltonian $H$, the unitarity condition is equivalent to the Hermiticity of $H$ with  respect to $\bbr~|~\kkt_{\rm inv}$ or alternatively its $\eta_{\rm inv}$-pseudo-Hermiticity. This implies that one may choose $\eta_{\rm inv}$ to be time-independent as well, i.e., set 
	\be
	\eta_{\rm inv}=\eta_0.
	\label{eq*}
	\end{equation}
It is not difficult to check that in this case the evolution operator 
$U(t,t_0)=e^{-i(t-t_0)H/\hbar}$ is {\em $\eta_0$-pseudo-unitary}, i.e.,
	\be
	U(t,t_0)^\sharp:=\eta_0^{-1}U(t,t_0)^\dagger\eta_0=U(t,t_0)^{-1},
	\label{p-unitary}
	\end{equation}
and Eqs.~(\ref{eq*}) and (\ref{eta-inv}) are identical. This observation also suggests a natural choice for $\eta_0$ in the general case where $H$ is time-dependent, namely to take $\eta_0$ to be a Hermitian, invertible, linear operator with respect to which the initial Hamiltonian
$H(t_0)$ is pseudo-Hermitian. In particular, if $H$ is diagonalizable and has a discrete spectrum, we may identify $\eta_0$ with $\tilde\eta_+$. In this way, the invariant inner product 
	\be
	\bbr~|~\kkt_{\rm inv}=\br~~|U^{-1\dagger}(t,t_0)\tilde\eta_+U(t,t_0)^{-1}|~~\kt
	\label{inv-inn-pro}
	\end{equation}
is positive-definite if $H$ has a real spectrum.

In summary, we have outlined a formulation of unitary quantum dynamics using a
time-dependent pseudo-Hermitian Hamiltonian $H(t)$ that acts in a fixed Hilbert space. This formulation relies on the idea of changing the original inner product $\br~|~\kt$ of the Hilbert space into an invariant inner product $\bbr~|~\kkt_{\rm inv}=\br~|\eta_{\rm inv}(t)|~\kt$ which is generally time-dependent. The latter is defined in an essentially unique way in terms of the inner product at the initial time $t_0$. For the case that $H(t)$ is diagonalizable and has a real spectrum, the natural choice for $\eta_0:=\eta_{\rm inv}(t_0)$ is a positive operator $\tilde\eta_+$ with respect to which $H(t_0)$ is pseudo-Hermitian. This choice is consistent with the fact that a time-independent Hamiltonian supports a unitary evolution (with respect to some inner product) if and only if it is pseudo-Hermitian.
In particular, if $H(t)$ is Hermitian, the invariant inner product $\bbr~|~\kkt_{\rm inv}$ reduces to the original inner product $\br~|~\kt$ on ${\cal H}$. Therefore, our treatment is a generalization of the ordinary unitary quantum mechanics with a fixed Hilbert space.

\section{Klein-Gordon Type Evolution Equations}

Consider a physical system with a linear evolution equation. Then it is well-known that the solution space of this equation is isomorphic as a vector space to the space of all possible initial data. In nonrelativistic quantum mechanics the evolution equation is the time-dependent Schr\"odinger equation (\ref{sch-eq}) which is first order in time. Therefore the solution space, which we identify with the physical Hilbert space, is isomorphic to the vector space of the initial state vectors. This vector space isomorphism may be promoted to a Hilbert space isomorphism, because the inner product of any two solutions is independent of time. Therefore, we may view the Hilbert space either as the space of initial conditions or the space of solutions of the Schr\"odinger equation. In the following we shall use this dual picture of the Hilbert space in the study of a class of linear evolution equations which are second order in time. 

Consider the evolution equations of the form 
	\be
	\ddot \psi + D \psi =0,
	\label{eq1}
	\end{equation}
where a dot denotes a time-derivative, $\psi$ belongs to a Hilbert space $\tilde{\cal H}$ with inner product $\br~|~\kt$, and $D:\tilde{\cal H}\to\tilde{\cal H}$ is a possibly time-dependent linear Hermitian operator. Because Eq.~(\ref{eq1}) involves a second order derivative with respect to time $t$, the dynamics is determined by two initial conditions $\psi(t_0)$ and $\dot\psi(t_0)$, or any two linearly independent linear combinations,
	\be
	u_0= a \psi(t_0)+b\dot\psi(t_0),~~~~
	v_0= c \psi(t_0)+d\dot\psi(t_0),
	\label{u-v}
	\end{equation}
of $\psi(t_0)$ and $\dot\psi(t_0)$. In Eqs.~(\ref{u-v}), $a,b,c$, and $d$ are the entries of an invertible complex $2\times 2$ matrix $g$. 

Now, suppose that the field equation (\ref{eq1}) describes the dynamics of a physical system. The states of the system are represented by the solutions of this equation, and the phase space is identified with the space of solutions of this equation modulo its symmetries. The space of solutions is isomorphic (as a vector space) to the space of the initial conditions. The latter has the vector space structure of 
	\[ {\cal H}:=\C^2\otimes\tilde{\cal H},\]
for an initial state vector may be represented as
	\be
	\Psi_0:=\left(\begin{array}{c} u_0\\ v_0\end{array}\right).
	\label{initial}
	\end{equation}
This observation suggests a two-component formulation of the dynamics of the system which makes its phase space structure transparent. In this formulation, the state vectors of the system belong to ${\cal H}$, and the evolution equation~(\ref{eq1}) takes the form of the Schr\"odinger equation $i\hbar\dot\Psi=H\Psi$, where $\Psi$ and $H$ are respectively the two-component state vector and the Hamiltonian:
	\bea
	\Psi&:=&\left(\begin{array}{c}
	\psi+i\lambda\dot\psi\\
	\psi-i\lambda\dot\psi\end{array}\right),
	\label{p-h}\\
	H&:=&\frac{\hbar}{2}\left(\begin{array}{cc}
	\lambda D+\lambda^{-1}&\lambda D-\lambda^{-1} \\
	-\lambda D+\lambda^{-1}&-\lambda D-\lambda^{-1}\end{array}\right),
	\label{p-h2}
	\eea
and $\lambda$ is an arbitrary nonzero real parameter which has the dimension of time.

The choice of the two-component state vector (\ref{p-h}) is obviously not unique. The general form of a two-component state vector is $\Psi_{g(t)}=g(t)\Psi$ where $g(t)$ is a time-dependent element of the general linear group $GL(2,\C)$. Eq.~(\ref{eq1}) is equivalent to the Schr\"odinger equation
$i\hbar\dot\Psi_{g(t)}=H_{g(t)}\Psi_{g(t)}$ with $H_{g(t)}$ given by
	\be
	H_{g(t)}=g(t) H g(t)^{-1}+i\hbar \dot g(t)g(t)^{-1}.
	\label{H-g}
	\end{equation}
The choice of $g(t)$ is completely arbitrary. As suggested by Eq.~(\ref{H-g}) this arbitrariness has its root in a nonphysical $GL(2,\C)$ gauge symmetry of the two-component formulation of the dynamics.\footnote{Sometimes one may use this gauge symmetry to simplify the analysis of the problem at hand. A good example is the two-component formulation of the Klein-Gordon equation in Bianchi type background spacetimes \cite{tjp}.} In the following we shall set $g(t)=1$. This is a partial gauge-fixing as the arbitrariness in the value of the parameter $\lambda$ is intact. In fact, one can show that changing $\lambda$ corresponds to gauge transformations associated with a $GL(1,\R)$
subgroup of $GL(2,\C)$, \cite{jpa-98}. As we shall see, our final results will be independent of $\lambda$.

The eigenvalue problem for the Hamiltonian~(\ref{p-h2}) may be easily solved. The eigenvalues $E_n$ and the corresponding eigenvectors $\Psi_n$ are given by
	\[E_n=\hbar\omega_n,~~~~~~~~~\Psi_n=\left(\begin{array}{c}
	\lambda^{-1}+\omega_n\\
	\lambda^{-1}-\omega_n\end{array}\right)\phi_n,\]
where $\omega_n$ and $\phi_n$ satisfy
	\be
	D\phi_n=\omega_n^2\phi_n.
	\label{eg-va-D}
	\end{equation}

Because $D$ is a Hermitian operator, its eigenvalues $\omega_n^2$ are real and its eigenvectors $\phi_n$ are orthogonal. This implies that the eigenvalues $E_n$ of $H$ are either real or come in complex-conjugate pairs; the spectrum is pseudo-real. In view of Theorem~I$\!$I$\!$I, this suggests that $H$ is pseudo-Hermitian. Note that $H$ is not Hermitian with respect to the $L^2$-inner product on ${\cal H}$, and that a loss of diagonalizability occurs if zero belongs to the spectrum of $D$. This is precisely the situation considered in Ref.~\cite{p7}. As shown in \cite{p7} this type of loss of diagonalizability does not violate the equivalence of the pseudo-reality of the spectrum and the pseudo-Hermiticity of the Hamiltonian. In fact the latter can be directly verified; a simple calculation shows that $H$ is $\sigma_3$-pseudo-Hermitian where $\sigma_3$ is the Pauli matrix $\sigma_3={\rm diag}(1,-1)$. Because $\sigma_3$ does not depend on time, it defines an invariant albeit indefinite pseudo-inner product, namely
	\be
	\bbr\Psi_1|\Psi_2\kkt_{\sigma_3}=\br\Psi_1|\sigma_3\Psi_2\kt=
	2i\lambda(\br\psi_1|\dot\psi_2\kt-\br\psi_2|\dot\psi_1\kt).
	\label{kg-prod}
	\end{equation}
Here and in what follows the two-component state vectors $\Psi_i$, with $i=1,2$, are related to the one-component state vectors $\psi_i$ according to (\ref{p-h}). The invariant inner product (\ref{kg-prod}) is known as the Klein-Gordon inner product.

Next, consider the special case where the spectrum of $D$ is positive, discrete, and nondegenerate. Then $H$ has a real, discrete, and nondegenerate spectrum, and as discussed in Section~2 it is Hermitian with respect to the positive-definite inner product $\bbr~|~\kkt_{\eta_+}$. In order to compute $\eta_+$, we first suppose that the spectral label $n$ takes nonnegative integer values, let
$\omega_n\in\R^+$, and express the eigenvalues and the eigenvectors of $H$ in the form
	\bea
	E_{\pm,n}&=&\pm \hbar\omega_n,
	\label{eg-va}\\
	\Psi_{\pm,n}&=&\left(\begin{array}{c}
	\lambda^{-1}\pm\omega_n\\
	\lambda^{-1}\mp\omega_n\end{array}\right)\phi_n,
	\label{eg-ve}
	\eea
We also assume without loss of generality that $\phi_n$ form a complete orthonormal set of eigenvectors of $D$, so that
	\be
	\br\phi_m|\phi_n\kt=\delta_{mn},~~~~~~\sum_n|\phi_n\kt\br\phi_n|=1,~~~~~~
	D=\sum_{n} \omega_n^2 |\phi_n\kt\br\phi_n|.
	\label{D=}
	\end{equation}
Now, we can calculate the eigenvectors $\Phi_{\pm,n}$ of $H^\dagger$ and the operator
	\be
	\eta_+=\sum_{n}( |\Phi_{+,n}\kt\br\Phi_{+,n}|+|\Phi_{-,n}\kt\br\Phi_{-,n}|).
	\label{eta+2}
	\end{equation}
This yields
	\bea
	\Phi_{\pm,n}&=&\frac{1}{4}\,
	\left(\begin{array}{c}
	\lambda\pm\omega_n^{-1}\\
	\lambda\mp\omega_n^{-1}\end{array}\right)\phi_n,
	\label{phi=}\\
	\eta_+&=&\frac{1}{8}\sum_{n}
	\left(\begin{array}{cc}
	\lambda^2+\omega_n^{-2}&\lambda^2-\omega_n^{-2}\\
	\lambda^2-\omega_n^{-2}&\lambda^2+\omega_n^{-2}\end{array}\right)|\phi_n\kt\br\phi_n|=
	\frac{1}{8}\left(\begin{array}{cc}
	\lambda^2+ D^{-1}&\lambda^2- D^{-1} \\
	\lambda^2-D^{-1}&\lambda^2+D^{-1}\end{array}\right),
	\label{eta+=}
	\eea
where we have made use of (\ref{D=}) and 
	\be
	D^{\gamma}=\sum_{n} \omega_n^{2\gamma} |\phi_n\kt\br\phi_n|,~~~~~\forall\gamma\in\R.
	\label{D-1=}
	\end{equation}
It is not difficult to check that indeed $\{\Psi_{\pm,n},
\Phi_{\pm,n}\}$ forms a complete biorthonormal system for ${\cal H}$.

Having obtained $\eta_+$ we can compute the inner product $\bbr~|~\kkt_{\eta_+}$ for any pair $\Psi_1$ and $\Psi_2$ of evolving two-component state vectors (\ref{p-h}). The resulting expression,
namely
	\be
	\bbr\Psi_1|\Psi_2\kkt_{\eta_+}=\br\Psi_1|\eta_+\Psi_2\kt=
	\frac{\lambda^2}{2}\,\left(\br\psi_1|\psi_2\kt+\br\dot\psi_1|D^{-1}|\dot\psi_2\kt\right),
	\label{+inn}
	\end{equation}
is surprisingly simple. One can check that indeed (\ref{+inn}) is a positive-definite inner product on ${\cal H}$ and that $H$ is Hermitian with respect to this inner product.

Now, suppose that $D$ does not depend on time. Then $\eta_+$ and the inner product $\bbr~|~\kkt_{\eta_+}$ are time-independent and as a consequence of Theorem~I$\!$I, $\bbr~|~\kkt_{\eta_+}$ is an invariant inner product. In particular, we can introduce 
	\be
	\cbr\psi_1,\psi_2\ckt_{\eta_+}:=\lambda^{-2}\bbr\Psi_1|\Psi_2\kkt_{\eta_+}=
	\frac{1}{2}\,\left(\br\psi_1|\psi_2\kt+\br\dot\psi_1|D^{-1}|\dot\psi_2\kt\right),
	\label{+inn-1}
	\end{equation}
which does not involve the arbitrary parameter $\lambda$ and therefore defines an invariant positive-definite inner product on the space of solutions of the original evolution equation~(\ref{eq1}). 

Recall that in quantum mechanics, it is the ratios of the inner products of the state vectors that enter in the calculation of the physical quantities. Therefore, the inner products (\ref{+inn}) and (\ref{+inn-1}) are physically equivalent. 

As we pointed out in Section~2, $\eta_+$ and the corresponding inner products $\bbr~|~\kkt_{\eta_+}$ and $\cbr~,~\ckt_{\eta_+}$ are not unique. The most general positive operator $\tilde\eta_+$ with respect to which the Hamiltonian~(\ref{p-h2}) is pseudo-Hermitian is given by Eq.~(\ref{eta-trans}) where $A$ has the general form
	\be
	A=\sum_n (a^+_{n} |\Psi_{+,n}\kt\br\Phi_{+,n}|+
	a^-_{n} |\Psi_{-,n}\kt\br\Phi_{-,n}|),
	\label{A=}
	\end{equation}
and $a^\pm_{n}$ are nonzero complex numbers. Substituting (\ref{eta+2}) and (\ref{A=}) in 
(\ref{eta-trans}), making use of the fact that $\{\Psi_{\pm,n},\Phi_{\pm,n}\}$ is a complete biorthonormal system, and employing Eqs.~(\ref{phi=}) and (\ref{D-1=}), we find
{\small	
	\bea
	\tilde\eta_+&=&\frac{1}{16}\sum_n\left(\begin{array}{cc}
	|a^+_n|^2(\lambda+\omega_n^{-1})^2+|a^-_n|^2(\lambda-\omega_n^{-1})^2 &
	(|a^+_n|^2+|a^-_n|^2)(\lambda^2-\omega_n^{-2})\\
	(|a^+_n|^2+|a^-_n|^2)(\lambda^2-\omega_n^{-2}) &
	|a^+_n|^2(\lambda-\omega_n^{-1})^2+|a^-_n|^2(\lambda+\omega_n^{-1})^2 
	\end{array}\right)|\phi_n\kt\br\phi_n|\nn\\
	&&\nn\\
	&=&\frac{1}{8}\left(\begin{array}{cc}
	L_+(\lambda^2+D^{-1})+2\lambda L_-D^{-1/2} & L_+(\lambda^2-D^{-1})\\
	L_+(\lambda^2-D^{-1}) & L_+(\lambda^2+D^{-1})-2\lambda L_-D^{-1/2}
	\end{array}\right),
	\label{tilde-eta}
	\eea
}%
where $L_\pm:\tilde{\cal H}\to\tilde{\cal H}$ are linear operators defined by
	\be
	L_\pm:=\frac{1}{2}\sum_n(|a_n^+|^2\pm|a_n^-|^2)|\phi_n\kt\br\phi_n|.
	\label{L}
	\end{equation}
As seen from this equation $L_\pm$ are Hermitian operators commuting with $D$, and $A_\pm:=L_+\pm L_-$ are positive operator.

Next, we compute the inner product $\bbr~|~\kkt_{\tilde\eta_+}$. In view of Eqs.~(\ref{p-h}), 
(\ref{tilde-eta}), and (\ref{D-1=}), we obtain after a rather lengthy calculation,
{\small
	\bea
	\bbr\Psi_1|\Psi_n\kkt_{\tilde\eta_+}&=&\br\Psi_1|\tilde\eta_+\Psi_2\kt\nn\\
	&=&
	\frac{\lambda^2}{2}\left[\br\psi_1|L_+|\psi_2\kt+\br\dot\psi_1|L_+D^{-1}|\dot\psi_2\kt
	+i(\br\psi_1|L_-D^{-1/2}|\dot\psi_2\kt-\br\dot\psi_1|L_-D^{-1/2}|\psi_2\kt)\right].\nn\\
	\label{tilde-inner}
	\eea
}%
The fact that the nonphysical parameter $\lambda$ just scales the inner product~(\ref{tilde-inner}) and therefore allows for the introduction of the positive-definite inner product,
	\bea
	\cbr\psi_1,\psi_2\ckt_{\tilde\eta_+}&:=&\lambda^{-1}\bbr\Psi_1|\Psi_2\kkt_{\tilde\eta_+}\nn\\
	&=&
	\frac{1}{2}\left[\br\psi_1|L_+|\psi_2\kt+\br\dot\psi_1|L_+D^{-1}|\dot\psi_2\kt
	+i(\br\psi_1|L_-D^{-1/2}|\dot\psi_2\kt-\br\dot\psi_1|L_-D^{-1/2}|\psi_2\kt)\right],\nn\\
	\label{xx}
	\eea
is very remarkable.\footnote{In fact $\lambda$ is present throughout the above calculation till the very last step where its contributions to various terms in $\bbr\Psi_1|\Psi_n\kkt_{\tilde\eta_+}$ cancel almost miraculously and only the trivial multiplicative factor $\lambda^2$ survives.}

Another important feature of the inner product (\ref{xx}) is that we can directly check its invariance by computing its time-derivative. A straightforward calculation shows that in view of Eq.~(\ref{eq1}) and the fact that $L_\pm$ commute with any power of $D$, the time-derivative of the right-hand side of (\ref{xx}) vanishes identically. 

For the case that the operator $D$ is time-independent, Eq.~(\ref{xx}) provides the general form of an invariant positive-definite inner product on the space of solutions of the Klein-Gordon type evolution equation~(\ref{eq1}). The operators $L_\pm$ appearing in Eq.~(\ref{xx}) are uniquely determined in terms of the arbitrary positive real numbers $|a^\pm_n|^2$ (equivalently the positive operators 
$A_\pm$). The following theorem summarizes our results for the case that $D$ is time-independent.
	\begin{itemize}
	\item[]{\bf Theorem V$\!$I:} Consider the evolution equation $\ddot\psi+D\psi=0$ where $D$ is a Hermitian operator acting in a Hilbert space $\tilde{\cal H}$ and has a real, positive, discrete, and nondegenerate spectrum. Then if $D$ does not depend on time, the general form of an invariant positive-definite inner product on the space of solutions of the evolution equation is given by 
	\be
	\cbr\psi_1,\psi_2\ckt:=\frac{1}{2}\left[\br\psi_1|L_+
	|\psi_2\kt+\br\dot\psi_1|L_+D^{-1}|\dot\psi_2\kt
	+i(\br\psi_1|L_-D^{-1/2}|\dot\psi_2\kt-\br\dot\psi_1|L_-D^{-1/2}|\psi_2\kt)\right],
	\label{thm6}
	\end{equation}
where $L_\pm$ are Hermitian operators acting in $\tilde{\cal H}$ such that
$A_\pm:=L_+\pm L_-$ are positive operators commuting with $D$.
	\end{itemize}
Furthermore, in light of Theorem~V, we have:
	\begin{itemize}
	\item[]{\bf Theorem V$\!$I$\!$I:} Let $D$ be as in Theorem~V$\!$I, but suppose that it depends on time. Then an invariant positive-definite inner product on the solution space of the evolution equation $\ddot\psi+D\psi=0$ that reduces to (\ref{thm6}) for time-independent $D$
has the form,
	\be
	\cbr \psi_1,\psi_2\ckt_{\rm inv}:=\left.\cbr\psi_1,\psi_2\ckt\right|_{t=t_0},
	\label{inv=KG}
	\end{equation}
where $\psi_i$, with $i=1,2$, are any two solutions, $t_0$ is the initial time, and $\cbr~,~\ckt$ is given by Eq.~(\ref{thm6}).
	\end{itemize}
As a final note of this section, we wish to emphasize that the restriction that the spectrum of $D$ be nondegenerate may be lifted without any reservations; Theorems~V$\!$I and V$\!$I$\!$I hold
for the cases that $D$ has degenerate eigenvalues.

\section{Classical Simple Harmonic Oscillator}

Consider the classical equation of motion for a simple harmonic oscillator of frequency $\omega$, i.e.,
	\be
	\ddot x+\omega^2 x=0.
	\label{sho}
	\end{equation}
This is clearly a special case of a Klein-Gordon type evolution equation~(\ref{eq1}). We can apply the results of Section~4, by setting $\psi=x, D=\omega^2, \tilde{\cal H}=\C$, and ${\cal H}=\C^2$.
As in this case $D$ is the operation of multiplication by the positive real number $\omega^2$, we have $n=0$, $\omega_0=\omega$, $D^{-1}=\omega^{-2}$, and $D^{-1/2}=\omega^{-1}$. We can also set $\phi_0=1$. Furthermore, the positive-definite inner product~(\ref{xx}) (alternatively (\ref{thm6})) takes the form
	\be
	\cbr x_1,x_2\ckt_{\tilde\eta_+}=\cbr x_1,x_2\ckt=
	\frac{1}{2}\left[L_+(x_1^*x_2+\omega^{-2}\dot x_1^*\dot x_2)
	+iL_-\omega^{-1}(x_1^*\dot x_2-\dot x_1^*x_2)\right],
	\label{xx-sho}
	\end{equation}
where $x_i$, with $i=1,2$, are any two complex-valued solutions of (\ref{sho}), and $L_\pm$ are
any pair of real numbers such that $A_\pm:=L_+\pm L_-$ are positive. According to Theorem~V$\!$I,
Eq.~(\ref{xx-sho}) yields the most general invariant positive-definite inner product on the space of (complex) solutions of Eq.~(\ref{sho}) provided that the frequency $\omega$ does not depend on time. 

It is instructive to compute the inner product of the basic complex solutions $\zeta_\epsilon:=e^{-i\epsilon\omega t}$ where $\epsilon=\pm$. The result is
	\be
	\cbr\zeta_{\epsilon'},\zeta_\epsilon\ckt_{\tilde\eta_+}=\delta_{\epsilon'\epsilon}\;A_\epsilon,
	\label{zz}
	\end{equation}
where $\epsilon,\epsilon'=\pm$. Eq.~(\ref{zz}) offers a clear demonstration of the invariance and positive-definiteness of the inner product $\cbr~~,~~\ckt_{\tilde\eta_+}$. Choosing $L_+=1$ and $L_-=0$, so that $A_\pm=1$, we have $\tilde\eta_+=\eta_+$ and
$\cbr\zeta_{\epsilon'},\zeta_\epsilon\ckt_{\eta_+}=\delta_{\epsilon'\epsilon}$. Hence, the basic solutions are orthonormal with respect to the inner product $\cbr~,~\ckt_{\eta_+}$. This is in contrast 
with the Klein-Gordon inner product (\ref{kg-prod}),
	\[ \cbr\zeta_{\epsilon'},\zeta_\epsilon\ckt_{\rm KG}:=
	2i\lambda(\br\zeta_{\epsilon'}|\dot\zeta_\epsilon\kt-\br\zeta_\epsilon|\dot\zeta_{\epsilon'}\kt)=
	4\lambda\;\epsilon\;\delta_{\epsilon'\epsilon},\]
which is clearly indefinite. The difference becomes even more drastic if we consider the basic real solutions $z_1=\sin(\omega t)$ and $z_2=\cos(\omega t)$ which have zero Klein-Gordon norm but  positive real norm in the inner product $\cbr~~,~~\ckt_{\tilde\eta_+}$ or $\cbr~~,~~\ckt_{\eta_+}$. 
	
If the frequency $\omega$ depends on time, the inner product $\cbr~~,~~\ckt_{\tilde\eta_+}$ and in particular $\cbr~~,~~\ckt_{\eta_+}$ fail to be invariant. In this case, one makes use of Theorem~V$\!$I$\!$I and obtains the following expression for a general invariant positive-definite inner product which is valid for both time-dependent and time-independent frequencies.
	\[ \cbr x_1,x_2\ckt_{\rm inv}=\left.\cbr x_1,x_2\ckt\right|_{t=t_0}=
	\left.\frac{1}{2}\left[L_+(x_1^*x_2+\omega^{-2}\dot x_1^*\dot x_2)
	+iL_-\omega^{-1}(x_1^*\dot x_2-\dot x_1^*x_2)\right]\right|_{t=t_0}.\]

\section{Klein-Gordon Equation}

Another special case of the evolution equation~(\ref{eq1}) is the Klein-Gordon equation
	\be
	-\ddot\psi(\vec x,t)+\nabla^2\psi(\vec x,t)=\mu^2\psi(\vec x,t),
	\label{kg}
	\end{equation}
where a dot means a derivative with respect to $x^0:=c\,t$, $c$ is the velocity of light, $\mu:=m\,c/\hbar$, and $m$ is the mass of the Klein-Gordon field $\psi:\R^{3+1}\to\C$.
The two-component formulation of the Klein-Gordon equation~(\ref{kg}) has been considered in the 1950's \cite{fv}. More recently, it has been used in the study of the relativistic geometric phases \cite{jpa-98,tjp}. A detailed textbook treatment is offered in \cite{G}. 

We can express Eq.~(\ref{kg}) in the form~(\ref{eq1}) by setting 
	\be
	D:=-\nabla^2+\mu^2.
	\label{D-KG}
	\end{equation} 
Obviously, $D$ is a positive Hermitian operator acting in $\tilde{\cal H}=L^2(\R^3)$. But it has a degenerate and continuous spectrum. Although we have stated our general results for the case that $D$ has a discrete and nondegenerate spectrum, we can check that they apply to this case. As we pointed out earlier, the degeneracy of the spectrum can be easily incorporated into our method, and we can treat the continuous spectrum of $D$ as the limit of the discrete spectrum corresponding to the approximation in which one identifies the space $\R^3$ with the volume of a cube of side $\ell$ as $\ell$ tends to infinity.\footnote{$D$ is essentially the Hamiltonian for a nonrelativistic free particle. Its eigenvectors do not belong to $L^2(\R^3)$. They are generalized eigenvectors describing scattering states.}

In this section we shall replace the spectral label $n$ by the vector $\vec k\in\R^3$. This is because the eigenvectors of $D$ and the corresponding eigenvalues are respectively given by
	\be
	\phi_{\vec k}(\vec x):=\br\vec x|\vec k\kt=(2\pi)^{-3/2}e^{i\vec k\cdot\vec x},~~~~~
	\omega^2_{\vec k}=k^2+\mu^2,
	\label{k}
	\end{equation}
where $k^2:=\vec k\cdot\vec k$. We can apply the results of Section~4 provided that we make the following changes
	\be
	n\to\vec k,~~~~~~~\sum_n\to \int d^3k,~~~~~~~\delta_{n'n}\to\delta(\vec k'-\vec k).
	\label{change}
	\end{equation}
For example, applying (\ref{change}) to (\ref{D=}), we obtain the orthonormality and completeness conditions for $\phi_{\vec k}$ and the spectral resolution of $D$, namely
	\be
	\br\vec k'|\vec k\kt=\delta^3(\vec k'-\vec k),~~~~~~~
	\int d^3k |\vec k\kt\br\vec k|=1,~~~~~~~
	D=\int d^3k (k^2+\mu^2)|\vec k\kt\br\vec k|.
	\label{D=kg}
	\end{equation}
The eigenvectors $\Psi_{\pm,\vec k}$ and the eigenvalues $E_{\pm,\vec k}$ of the Hamiltonian~(\ref{p-h2}) are respectively given by Eqs.~(\ref{eg-ve}) and (\ref{eg-va}) with $n$ replaced with $\vec k$. Similarly the eigenvectors $\Phi_{\pm,\vec k}$ of $H^\dagger$, which together with $\Psi_{\pm,\vec k}$ form a complete biorthonormal system, are obtained 
by setting $n=\vec k$ in Eq.~(\ref{phi=}). The biorthonormality and completeness 
conditions~(\ref{bi}) become
	\be
	\br\Psi_{\epsilon',\vec k'}|\Phi_{\epsilon,\vec k}\kt=
	\delta_{\epsilon'\epsilon}\delta^3(\vec k'-\vec k),~~~~~~~~
	\sum_\epsilon\int d^3k\: |\Psi_{\epsilon,\vec k}\kt\br\Phi_{\epsilon,\vec k}|=1,
	\label{bi-kg}
	\end{equation}
where $\epsilon,\epsilon'=\pm$. 

One can repeat the analysis of \cite{p1,p2,p3,p4} for this case and show that because the spectrum of $H$ is real, it must be Hermitian with respect to a positive-definite inner product, namely
	\be
	\eta_+=\sum_\epsilon\int d^3k\: |\Phi_{\epsilon,\vec k}\kt\br\Phi_{\epsilon,\vec k}|
	=\frac{1}{8}\int d^3k
	\left(\begin{array}{cc}
	\lambda^2+ (k^2+\mu^2)^{-1}&\lambda^2- (k^2+\mu^2)^{-1} \\
	\lambda^2-(k^2+\mu^2)^{-1}&\lambda^2+ (k^2+\mu^2)^{-1}
	\end{array}\right)|\vec k\kt\br\vec k|.
	\label{eta+kg1}
	\end{equation}
In view of Eqs.~(\ref{bi-kg}), we also have
	\[ D^{-1}=\int d^3k (k^2+\mu^2)^{-1}|\vec k\kt\br\vec k|.\]
Using this identity and the second equation in (\ref{D=kg}), we can expression (\ref{eta+kg1}) in the form
	\be
	\eta_+=\frac{1}{8}\left(\begin{array}{cc}
	\lambda^2+ D^{-1}&\lambda^2- D^{-1} \\
	\lambda^2-D^{-1}&\lambda^2+D^{-1}\end{array}\right),
	\label{eta-KG}
	\end{equation}
which coincides with the last equation in (\ref{eta+=}). This in turns means that 
expressions~(\ref{+inn}) and (\ref{+inn-1}) for the invariant positive-definite inner products $\bbr~|~\kkt_{\eta_+}$ and $\cbr~,~\ckt_{\eta_+}$ are still valid. Similarly, Eq.~(\ref{thm6}) yields the most general invariant positive-definite inner product on the space of solutions of the Klein-Gordon equation~(\ref{kg}), where now $L_\pm$ are Hermitian operators acting in $L^2(\R^3)$ and 
$A_\pm=L_+\pm L_-$ are positive operators commuting with $D=-\nabla^2+\mu^2$ or alternatively with the Laplacian $\nabla^2$. In view of (\ref{L}), we have
	\bea
	A_\pm&=&\int d^3k\: \alpha_\pm(\vec k) |\vec k\kt\br\vec k|,
	\label{A-kg}\\
	L_\pm&=&\frac{1}{2}\,\int d^3k\: [\alpha_+(\vec k)\pm \alpha_-(\vec k)]|\vec k\kt\br\vec k|,
	\label{L-kg}
	\eea
where $\alpha_\pm(\vec k)$ are positive real coefficients.

Next, we compute the inner product of the basic (free particle) solutions:
	\be
	\psi_{\epsilon,\vec k}=N_{\epsilon,\vec k}\; e^{-i\epsilon\omega_{\vec k} x^0}\phi_{\vec k},
	\label{basic}
	\end{equation}
where $N_{\epsilon,\vec k}$ are normalization constants. Using Eqs.~(\ref{thm6}), (\ref{k}), (\ref{D=kg}) and (\ref{L-kg}), we find after a straightforward calculation
	\bea
	\cbr\psi_{\epsilon',\vec k'},\psi_{\epsilon,\vec k}\ckt&=&
	\frac{1}{4}\,N_{\epsilon',\vec k}^*N_{\epsilon,\vec k}
	\left[(1+\epsilon\epsilon')(\alpha_++\alpha_-)+
	(\epsilon+\epsilon')(\alpha_+-\alpha_-)\right]
	e^{i(\epsilon'-\epsilon)\omega_{\vec k}x^0}\delta^3(\vec k'-\vec k)\nn\\
	&=&\alpha_\epsilon |N_{\epsilon,\vec k}|^2\;\delta_{\epsilon'\epsilon}\;
	\delta^3(\vec k'-\vec k),
	\label{inn-basic-kg}
	\eea
where we have used the abbreviation $\alpha_\pm$ for 
$\alpha_\pm(\vec k)$. As both $\alpha_\pm$ are positive real numbers and the right-hand side of 
Eq.~(\ref{inn-basic-kg}) does not involve $x^0$, this equation provides an explicit demonstration of the invariance and positive-definiteness of the inner product $\cbr~,~\ckt$. 

Having obtained the inner product for the basic solutions, we can compute the inner product for any two solutions:
	\be
	\psi_i=\sum_\epsilon\int d^3k\: c_i(\epsilon,\vec k)\psi_{\epsilon,\vec k},
	\label{general}
	\end{equation}
where $i=1,2$ and $c_i(\epsilon,\vec k)$ are complex coefficients. In view of
Eqs.~(\ref{inn-basic-kg}) and (\ref{general}) and the fact that $\cbr~,~\ckt$ is a Hermitian sesquilinear form, we have
	\be
	\cbr\psi_1,\psi_2\ckt=\sum_\epsilon\int d^3k\: \alpha_\epsilon(\vec k)
	|N_{\epsilon,\vec k}|^2 c_1^*(\epsilon,\vec k)c_2(\epsilon,\vec k).
	\label{gen-inn}
	\end{equation}

Next, we wish to recall that one of the appealing properties of the Klein-Gordon inner product is that it is relativistically invariant. The class of the invariant positive-definite inner products that we have constructed above also include relativistically invariant members. These correspond to the choices for
$\alpha_\pm$ that make the right-hand side of (\ref{gen-inn}) a Lorentz scalar. Supposing that the 
basic solutions~(\ref{basic}) are scalar, we see that the normalization constants $N_{\pm,\vec k}$ must also be scalar. On the other hand, we know that $d^3k/\omega_{\vec k}$ is a relativistically invariant measure \cite{weinberg}. Hence in view of (\ref{general}) and the fact that the solutions $\psi_i$ are scalars, we infer that $c_i$ obey the same Lorentz transformation rule as 
$\omega_{\vec k}^{-1}$. This in turn implies that in order for the inner product (\ref{gen-inn}) to be scalar, $\alpha_\pm(\vec k)$ must transforms as $\omega_{\vec k}$. In particular, we have
	\be
	\alpha_\pm(\vec k)=\mu^{-1}\omega_{\vec k}\; a_\pm,
	\label{condi}
	\end{equation}
where $a_\pm$ are dimensionless positive real scalars (numbers). Under the condition~(\ref{condi}), $\cbr~,~\ckt$ is not only an invariant and positive-definite inner product, but it is relativistically invariant as well.

Substituting (\ref{condi}) in (\ref{L-kg}) and using (\ref{D=kg}), we find
	\be
	L_\pm=\frac{1}{2\mu}\,(a_+\pm a_-)\, D^{1/2}.
	\label{Lpm=}
	\end{equation}
This in turn implies that the general form of the inner product (\ref{thm6}) that is relativistically invariant 
is given by
	\be
	\cbr\psi_1,\psi_2\ckt_{\rm r.i.}:=
	\frac{1}{4\mu}\left[(a_++a_-)(\br\psi_1|D^{1/2}|\psi_2\kt+\br\dot\psi_1|D^{-1/2}|\dot\psi_2\kt)
	+i(a_+-a_-)(\br\psi_1|\dot\psi_2\kt-\br\dot\psi_1|\psi_2\kt)\right].
	\label{thm6-2}
	\end{equation}
As seen from this equation, for the Klein-Gordon fields, the set of invariant positive-definite inner products that are relativistically invariant form a two parameter family of all the invariant positive-definite inner products. 

Because the Klein-Gordon equation is homogeneous, two solutions that differ by
a multiplicative constant are physically equivalent. This means that one can always absorb the first parameter, namely $a_++a_-$ which is positive, in the definition of the fields. Alternatively, one may
obtains the physically distinct inner products~(\ref{thm6-2}) by fixing the value of $a_++a_-$. This leads to the following theorem.
	\begin{itemize}
	\item[] {\bf Theorem I$\!$I$\!$X:} There is a one-parameter family of physically distinct, invariant, positive-definite, and relativistically invariant inner products on the space of solutions of the Klein-Gordon equation~(\ref{kg}) which are labeled by the elements of the open unit interval, $(-1,1)$. Specifically, such an inner product has the general form
	 \be
	\cbr\psi_1,\psi_2\ckt_{\rm r.i.}:=
	\frac{1}{2\mu}\left[\br\psi_1|D^{1/2}|\psi_2\kt+\br\dot\psi_1|D^{-1/2}|\dot\psi_2\kt
	+ia(\br\psi_1|\dot\psi_2\kt-\br\dot\psi_1|\psi_2\kt)\right],
	\label{thm6-3}
	\end{equation}
where $a\in(-1,1)$.
	\item[] {\bf Proof:} Setting $a_++a_-=2$, introducing $a:=(a_+-a_-)/2$, and using 
(\ref{thm6-2}), we obtain (\ref{thm6-3}) and the condition that $|a|<1$.~~~$\square$
	\end{itemize}

Next, we wish to address the problem of the nonrelativistic limit of the inner products (\ref{thm6}) and specifically (\ref{thm6-2}). In order to do this we take two Klein-Gordon fields $\psi_i$ and let
	\be
	\chi_i(x^0,\vec x):=e^{-i\mu x^0}\psi_i(x^0,\vec x).
	\label{eq01}
	\end{equation}
In the nonrelativistic limit where $c\to\infty$, $\chi_i$ may be shown to satisfy the nonrelativistic free Schr\"odinger equation \cite{G}, i.e., $\dot\chi_i=i\nabla^2\chi/(2\mu)$. Using this equation together with (\ref{eq01}) and doing the necessary algebra, one can show that the nonrelativistic limit of the inner product (\ref{thm6}) is 
	\be
	\cbr\psi_1,\psi_2\ckt\approx \br\chi_1|(L_++L_-)|\chi_2\kt=\br\chi_1|A_+|\chi_2\kt=
	\br\psi_1|A_+|\psi_2\kt.
	\label{thm6-nr}
	\end{equation}
In particular, if we demand relativistic invariance, i.e., enforce Eq.~(\ref{Lpm=}), we have $A_+=a_+D^{1/2}/\mu$. But in the nonrelativistic limit, $D^{1/2}\psi_2\approx \mu\psi_2$. Therefore, the nonrelativistic limit of the inner product (\ref{thm6-2}) is
	\be
	\cbr\psi_1,\psi_2\ckt_{\rm r.i.}\approx a_+\br\psi_1|\psi_2\kt.
	\label{thm6-2-nr}
	\end{equation}
Again we can absorb $a_+$ in the definition of $\psi_i$ and find that
	\be
	\cbr\psi_1,\psi_2\ckt_{\rm r.i.}\approx \br\psi_1|\psi_2\kt.
	\label{thm6-3-nr}
	\end{equation}
This equation shows that the nonrelativistic limit of the inner product (\ref{thm6-3}) is the $L^2$ inner product of nonrelativistic quantum mechanics. Therefore, besides its invariance, positive-definiteness, and relativistic invariance, the inner product~(\ref{thm6-3}) also has the correct nonrelativistic limit.
	
Finally, we wish to compare our results with those of Woodard \cite{woodard}. Woodard's inner product, in our notation and conventions, has the form
	\be
	(\psi_1,\psi_2)_{\rm W}=i\mu^{-1} (\br\psi_1^+|\dot\psi_2^+\kt-\br\psi_1^-|\dot\psi_2^-\kt),
	\label{wood}
	\end{equation}
where $\psi^\pm_i$ is the $\pm$ energy part of $\psi_i$. We can easily compute the Woodard inner product of two basic solutions~(\ref{basic}). The result is
	\be
	(\psi_{\epsilon',\vec k},\psi_{\epsilon,\vec k})_{\rm W}=\mu^{-1}
	\omega_{\vec k}\;|N_{\epsilon,\vec k}|^2\,\delta_{\epsilon'\epsilon}\;\delta^3(\vec k'-\vec k).
	\label{wood2}
	\end{equation}
Comparing this expression with (\ref{inn-basic-kg}), we see that Woodard's inner product corresponds to setting $\alpha_\pm=\mu^{-1}\omega_{\vec k}$. In view of (\ref{condi}) and (\ref{Lpm=}), this implies that Woodard's inner product is a special case of the relativistically invariant inner products (\ref{thm6-3}) corresponding to choice $a=0$, i.e.,
	\be
	(\psi_1,\psi_2)_{\rm W}=
	\frac{1}{2\mu}\left[\br\psi_1|D^{1/2}|\psi_2\kt+\br\dot\psi_1|D^{-1/2}|\dot\psi_2\kt\right].
	\label{wood3}
	\end{equation}
Note also that although expressions (\ref{wood3}) and (\ref{wood}) are equivalent, the latter does not involve the explicit splitting of $\psi_i$ into positive and negative energy parts.

\section{Minisuperspace Wheeler-DeWitt Equation}

Consider the Wheeler-DeWitt equation for a FRW model coupled to a massive real scalar field 
$\varphi$ of mass $m$,
	\be
	\left[ -\frac{\partial^2}{\partial\alpha^2}+\frac{\partial^2}{\partial\varphi^2}+
	\kappa\,e^{4\alpha}-m^2\,e^{6\alpha}\varphi^2\right]\,\psi(\alpha,\varphi)=0,
	\label{wdw}
	\end{equation}
where $\alpha:=\ln a$, $a$ is the scale factor, $\kappa=-1,0,1$ determines whether the FRW model 
describes an open, flat, or closed universe, respectively, and we have chosen a particularly simple factor ordering and the natural units, \cite{page,wiltshire}. The Wheeler-DeWitt equation~(\ref{wdw}) 
also belongs to the Klein-Gordon type equations studied in Section~4. We can write it in the form~(\ref{eq1}), if we identify $\alpha$ with the time variable and let
	\be
	D:=-\frac{\partial^2}{\partial\varphi^2}+m^2\, e^{6\alpha}\varphi^2-\kappa\,e^{4\alpha}.
	\label{D=wdw}
	\end{equation}
This operator is essentially the Hamiltonian operator for a time-dependent simple harmonic oscillator. It acts in the Hilbert space $L^2(\R)$ and has a nondegenerate discrete spectrum. Therefore, we can directly apply the results of Section~4. 

We can readily solve the eigenvalue equation~(\ref{eg-va-D}) for $D$. This yields \cite{jmp-98}
	\bea
	\omega_n&=&m\,e^{3\alpha}(2n+1)-\kappa\,e^{4\alpha},
	\label{wdw-eg-va}\\
	\phi_{n}&:=&\br\varphi|n\kt=N_n H_n(m^{1/2}e^{3\alpha/2}\varphi)\,
	e^{-m\,e^{3\alpha}\varphi^2/2},
	\label{wdw-eg-ve}
	\eea
where $n=0,1,2,\cdots$, $H_n$ are Hermite polynomials, and $N_n:=[m\,e^{3\alpha}/(\pi 2^{2n}{n!}^2)]^{1/4}$ are normalization constants. 

As seen from Eq.~(\ref{wdw-eg-va}), $D$ has a positive real spectrum for the open and flat universes
where $\kappa=-1,0$ and a nonpositive spectrum for sufficiently large values of the scale factor (namely $a\geq m$) for the closed universe where $\kappa=1$. For the open and flat universes and for $a<m$ in case of the closed universe, Eq.~(\ref{thm6}) together with 
	\[D^\gamma=\sum_n [m\,e^{3\alpha}(2n+1)-\kappa\,e^{4\alpha}]^\gamma |n\kt\br n|,~~~~
	\forall\gamma\in\R,\]
yield a positive-definite inner product on the solution space of (\ref{wdw}). However, as $D$ depends on $\alpha$, this inner product is not invariant. The situation is an infinite-dimensional analog of the time-dependent simple harmonic oscillator considered in Section~5. The most general invariant positive-definite inner product is given by Eq.~(\ref{inv=KG}) of Theorem~V$\!$I$\!$I. In fact, this inner product may also be used for the case of the closed universe for all $\alpha\in\R$ provided that one considers evolutions for which the initial value $a_0$ of the scale factor satisfies $a_0<m$.
The invariant inner product Eq.~(\ref{inv=KG}) involves the operators $L_\pm$ that according to Eq.~(\ref{L}) are determined in terms of two arbitrary sequences $\{ |a_n^\pm|^2\}$ of positive real numbers. 

The following theorem summarizes the above arguments.
	\begin{itemize}
	\item[] {\bf Theorem I$\!$X:} For the open and flat FRW models, to each choice of the initial scale factor $a_0$ there corresponds a countably infinite family of invariant positive-definite inner products on the space of solutions of the Wheeler-DeWitt equation~(\ref{wdw}). For the closed FRW model the same holds for all $a_0<m$.
	\end{itemize}
Perhaps, the simplest choice for the operators $L_\pm$ is $L_+=1$ and $L_-=0$. Substituting these equation in (\ref{inv=KG}) and making use of (\ref{thm6}) we find
	\be
	\cbr\psi_1,\psi_2\ckt=\left.\frac{1}{2}\:(\br\psi_1|\psi_2\kt+\br\dot\psi|D^{-1}|\dot\psi_2\kt)
	\right|_{a=a_0}
	\label{wdw-simple}
	\end{equation}

\section{Conclusion}

In this paper we have given a complete solution of the problem of determining the most general invariant positive-definite inner product on the space of solutions of a Klein-Gordon type field equation. Our solution relies on the following basic observations.
	\begin{enumerate}
	\item The two-component formulation of the field equation provides a natural framework to address the problem;
	\item The effective Hamiltonian appearing in the two-component form of the field equation is a pseudo-Hermitian Hamiltonian with a real spectrum, i.e., it is quasi-Hermitian;
	\item Every quasi-Hermitian Hamiltonian is pseudo-Hermitian with respect to a positive operator
$\eta_+$ which can be explicitly constructed;
	\item The Hermiticity of the Hamiltonian does not ensure the unitarity of the evolution for a Hilbert space that has a time-dependent inner product.
	\end{enumerate}

We started our analysis by addressing the issue of characterizing the invariant inner products that support a unitary evolution for a time-dependent pseudo-Hermitian operator. This problem is related to the apparently unexplored difficulties of formulating a unitary quantum mechanics in a time-dependent Hilbert space.
We then considered the Klein-Gordon type equations (\ref{eq1}) and showed how the machinery of the theory of pseudo-Hermitian Hamiltonians could be used to construct the most general invariant positive-definite inner product on the solution space of such equations. In particular, we explored the instructive example of a classical simple harmonic oscillator with both time-independent and time-dependent frequencies. We then applied our results to the free Klein-Gordon equation in 3+1 dimensions. We constructed the most general invariant positive-definite inner product for the Klein-Gordon fields and obtained the physically distinct, invariant positive-definite inner products that are also relativistically invariant. We explored the nonrelativistic limit of these inner products and showed that in this limit they tend to the $L^2$ inner product of the nonrelativistic quantum mechanics. We also compared our results with those of Woodard \cite{woodard} and found that as we expected Woodard's inner product is a special case of ours. Finally, we considered the application of our method to the Wheeler-DeWitt equation for the FRW-massive-real-scalar-field models that are of interest in the context of inflationary cosmology. For these models we showed that there is a countably infinite set of invariant positive-definite inner products provided that the universe is open or flat or that we take the initial scale factor to be less than the mass of the scalar field (in natural units). 

We conclude this paper with the following remarks.
	\begin{itemize}
	\item[--] Our results are obtained regardless of any qualitative arguments. Therefore, they have the advantage of providing a framework for imposing various physical restrictions to identify the `most appropriate' inner product.
	\item[--] Each choice of the invariant positive-definite inner product corresponds to a Hilbert space structure on the solution space of the field equation. This in turn allows for introducing the observables of the theory as the Hermitian operators acting in the Hilbert space. Furthermore, one can address a variety of physical problems such as the wave-packet dynamics, semi-classical evolutions, etc.
	\item[--] In our analysis we did not pay attention to the technical issues such as the domain of the operators. For the specific applications that we considered here these issues turn out not to be  important. Yet for more general situations one would need a more rigorous treatment.
	\item[--] As shown in \cite{jpa-98,tjp}, the two-component form of the Klein-Gordon equation may be easily generalized to arbitrary (possibly non stationary) curved backgrounds. Therefore, the method proposed in this article has a wider domain of application than the models considered here. We leave a more comprehensive study of these applications in particular in connection with quantum cosmology for a future publication.
	\end{itemize}

\section*{Acknowledgment}

This work has been supported by the Turkish Academy of Sciences in the framework of the Young Researcher Award Program (EA-T$\ddot{\rm U}$BA-GEB$\dot{\rm I}$P/2001-1-1).

\ed
\newpage
{\small
\begin{thebibliography}{99}
\bibitem{kuchar} K.\ Kuch\'ar, in {\em Proceedings of the 4th Canadian Conference on Relativity and Relativistic Astrophysics}, edited by G.\ Kunstatter, D.\ Vincent, and J.\ Williams (World Scientific, Singapore, 1992)
\bibitem{isham} C.\ J.\ Isham, in {\em Integrable Systems' Quantum Groups, and Quantum Field Theories,} edited by L.\ A.\ Ibort and M.\ A.\ Rodriguez (Kluwer, Dordrecht, 1993).
\bibitem{wiltshire} D.\ L.\ Wiltshire, in Cosmology: The Physics of the Universe, edited by B.\ Robson, N.\ Visvanathan, and W.\ S.\ Woolcock (World Scientific, Singapore, 1996).
\bibitem{carlip} S.~Carlip, Rep.\ Prog.\ Phys.\ {\bf 64}, 885 (2001).
\bibitem{dereli} T.\ Dereli, M.\ \"Ondar, and R.\ W.\ Tucker, Phys.\ Lett.\ B {\bf 324}, 134 (1994).
\bibitem{bryce-1} B.\ S.\ DeWitt, Phys.\ Rev.\ {\bf 160}, 1113 (1967).
\bibitem{vilenkin} A.\ Vilenkin, Phys.\ Rev.\ D {\bf 39}, 1116 (1989).
\bibitem{wald} R.\ M.\ Wald, Phys.\ Rev.\ D {\bf 48}, R2377 (1993).
\bibitem{page-w} D.\ N.\ Page and W.\ K\ Wootters, Phys.\ Rev.\ D {\bf 27}, 2885 (1983).
\bibitem{page} D.\ N.\ Page, in Gravitation: A Banff Summer Institute, edited by R.\ Mann and P.\ Wesson (World Scientific, Singapore, 1991).
\bibitem{woodard} P.\ P.\ Woodard, Class.\ Quantum.\ Grav.\ {\bf 10}, 483 (1993).
\bibitem{marolf} D.\ Marolf, Class.\ Quantum Grav.\ {\bf 12}, 1199 (1995); ibid arXiv: gr-qc/00011112;\\
A.~Ashtekar, J.~Lewandowski, D.~Marolf, J.~Mour\~ao, and T.~Thiemann, J.~Math.\ Phys.\ {\bf 36}, 6456 (1995).
\bibitem{gp} R.\ Gambini and R.\ A.\ Porto, Phys.\ Rev.\ D, {\bf 63}, 105014 (2001).
\bibitem{p1} A.~Mostafazadeh, J.\ Math.\ Phys., {\bf 43}, 205 (2002).
\bibitem{p2} A.~Mostafazadeh, J.\ Math.\ Phys., {\bf 43}, 2814 (2002).
\bibitem{p3} A.~Mostafazadeh, J.\ Math.\ Phys., {\bf 43}, 3944 (2002).
\bibitem{p4} A.~Mostafazadeh, Nucl.\ Phys.\ B {\bf 640}, 419 (2002).
\bibitem{p8} A.~Mostafazadeh, `Pseudo-Hermiticity and Generalized $PT$- and $CPT$-Symmetries,'
J.~Math.\ Phys., to appear (arXiv: math-ph/0209018).
\bibitem{kato} T.\ Kato, {\em Perturbation Theory for Linear Operators} (Springer, Berlin, 1995).
\bibitem{quasi} F.\ G.\ Scholtz, H.\ B.\ Geyer, and F.\ J.\ W.\ Hahne, Ann.\ Phys.\ {\bf 213}, 74 (1992).
\bibitem{ba} J.\ Bogn\'ar, {\em Indefinite Inner Product Spaces} (Springer, Berlin, 1974);
T.\ Ya.\ Azizov and I.\ S.\ Iokhvidov, {\em Linear Operators in Spaces with Indefinite Metric} (Wiley, Chichester, 1989).
\bibitem{pauli} W.\ Pauli, Rev.\  Mod.\ Phys., {\bf 15}, 175 (1943).
\bibitem{dirac} P.\ A.\ M.\ Dirac, Proc.\ Roy.\ Soc.\ London A {\bf 180}, 1 (1942).
\bibitem{p7} A.~Mostafazadeh, J.\ Math.\ Phys., {\bf 43}, 6343 (2002).
\bibitem{pra-97a} A.~Mostafazadeh, Phys.\ Rev.\ A {\bf 55}, 1653 (1997);\\
A. Mostafazadeh, J.\ Math.\ Phys., {\bf 40}, 3311 (1999).
\bibitem{nova} A.~Mostafazadeh, {\em Dynamical Invariants, Adiabatic Approximation, and the Geometric Phase} (Nova Science Publishers, New York, 2001).
\bibitem{tjp} A. Mostafazadeh, Turkish J. of Physics 24, 411 (2000).
\bibitem{jpa-98} A.~Mostafazadeh, J.\ Phys.\ A: Math.\ Gen., {\bf 31}, 7827 (1998).
\bibitem{fv} H.\ Feshbach and F.\ Villars, Rev.\  Mod.\ Phys., {\bf 30}, 24 (1958).
\bibitem{G} W.\ Greiner, {\em Relativistic Quantum Mechanics} (Springer, Berlin, 1994).
\bibitem{weinberg} S.~Weinberg, {\em The Quantum Theory of Fields}, Vol.~I (Cambridge University Press, Cambridge, 1995).
\bibitem{jmp-98} A. Mostafazadeh, J.\ Math.\ Phys., {\bf 39}, 4499 (1998).
\end{thebibliography} 
}
